\documentclass[prd,aps,nofootinbib,preprintnumbers,twocolumn]{revtex4}
\usepackage{amsmath,amssymb}
        \usepackage{amssymb}
        \usepackage{graphicx} %package para a insero de imagens
        \usepackage{rotating} %package que permite a incluso de tabelas "deitadas"
       \usepackage{multirow} %expanso de clulas por vrias linhas
        \usepackage{amsfonts}
        \usepackage{amsmath}
        \usepackage{subfigure}

\def\bq{\begin{quote}}
\def\eq{\end{quote}}

\begin{document}
%%%%%%%%%%%%%%%%%%%%%%%%%%%%%%%%%%%%%%%%%%%%%%%%%%%%%%%%%%%%%%%%%
\title{Magnetisation dynamics in exchange coupled spring systems with perpendicular anisotropy.}
\author{Pedro M. S. Monteiro\footnote{Electronic address: pedrmonteiro@gmail.com}, D. S. Schmool\footnote{Electronic address: dschmool@fc.up.pt}}
\affiliation{Departamento de F\'{i}sica and IFIMUP, Universidade do Porto,
Rua do Campo Alegre 687, 4169-007 Porto, Portugal}

\begin{abstract}

Magnetisation dynamics in exchange spring magnets have been studied using simulations of the FePt/Fe bilayer system. The FePt hard layer exhibits strong perpendicular magnetocrystalline anisotropy, while the soft (Fe) layer has negligible magnetocrystalline anisotropy. The variation of the local spin orientation in the Fe layer is determined by the competition of the exchange coupling interaction with the hard layer and the magnetostatic energy which favours in-plane magnetisation. Dynamics were studied by monitoring the response of the Fe layer magnetisation after the abrupt application of a magnetic field which causes the systems to realign via precessional motion. This precessional motion allows us to obtain the frequency spectrum and hence examine the dynamical magnetisation motion. Since the rotation of the spins in the soft layer does not have a well defined magnetic anisotropy, the system does not present the usual frequency  field characteristics for a thin film. Additionally we obtain multi-peaked resonance spectra for the application of magnetic fields perpendicular to the film plane, though we discount the existence of spin wave modes and propose that this arises due to variations in the local effective field across the Fe layer. The dynamic response is only considered in the Fe layer, with the FePt layer held fixed in the perpendicular orientation.

\vskip 1cm

\end{abstract}

\preprint{arXiv:0911.4137}

\maketitle

\section{Introduction}\label{introduction}

Increasing attention has been paid to the so-called exchange-spring systems, which couple the hard and soft magnetic properties of the two magnetic components via the exchange interaction between them. Recent studies have indicated that the magnetic energy product can be theoretically as high as $120$ MGOe \cite{1}. The exchange spring system is in general characterised by the reversible dynamics in the soft component arising from the competition between the magnetic anisotropies in the two magnetic materials. Goto \textit{et al.} \cite{2} were the first to consider such systems based on the assumption that the exchange and anisotropy energies are large enough to overcome the magnetostatic energy of the system. The specific spin configuration for such a system will depend on the relative strengths and directions of the anisotropies, as well as the exchange interaction between the magnetic layers (phases) and the orientation of any applied magnetic field. The spin configuration is then determined by the energy minimisation, taking into account the various magnetic contributions in the system as a whole. Kneller and Hawig \cite{3} applied these principles to the construction of a new class of permanent magnetic materials which are referred to as the exchange  springs. These authors analysed the reversible component of the hysteresis loop. Fullerton et al \cite{4} considered layered systems of Sm-Co/Fe and Sm-Co/Co, performing simulations based on a $1$D model. Bowden \textit{et al.} \cite{5} have made extensions to Goto's $1$D model for discrete systems and have obtained a relation between the bending and exchange fields. Bowden \textit{et al.} \cite{1} have also performed numerical simulations of $1$, $2$ and $3$D systems, using point defects in the chain to determine the behaviour of the system and predict a collapse of the exchange spring system.
The model of Asti \textit{et al.} \cite{6} considers a FePt (hard) magnetic layer with perpendicular anisotropy coupled to a Fe (soft) magnetic layer. Among the various considerations, the authors have studied the equilibrium phase diagram as a function of the thicknesses of the two magnetic layers, predicting a transition from a rigid magnet (RM) to an exchange spring (ES). The rigid magnet regime concerns the state where the soft magnetic layer magnetisation strictly follows that of the hard layer. The exchange spring effect is manifest in the local variation of the effective field. In the bilayered system that we are considering, the FePt layer has dominant perpendicular magnetocrystalline anisotropy, while the Fe layer has dominant in-plane anisotropy due to shape effects. As the Fe layer thickness increases, the local field reflects the varying influences of the two contributions and the local spin orientation is seen to rotate from the perpendicular direction at the FePt/Fe.

Ferromagnetic resonance can provide explicit information regarding interlayer interactions in multilayered structures and has been successfully applied to many systems \cite{w} and \cite{ww}. 
There are very few papers that deal explicitly with FMR in exchange spring systems, so it is therefore a new challenge that needs to be studied. Grimsditch et al \cite{Grim} studied experimentally the magnon frequency dependence of a Sm-Co/Fe system suing Brillouin scattering and found a good agreement with the theory developed for the one-dimensional model. Also, Pechan et al \cite{Pechan} investigated the frequency dependence of the ferromagnetic modes as a function of the in-plane angle.

Crew and Stamps \cite{www} have simulated a bilayered system of CoPt/Co, to study the resonance frequency spectrum and anisotropy for a particular applied field direction. However, this paper offers a more detailed description of the dynamical behaviour of these kind of systems, contributing to a better understanding. 

Experimentally, FePt/Fe system has been studied by Casoli \textit{et al.} \cite{7}, who have performed measurements of the hysteretic behaviour, confirming the transition from rigid magnet to exchange spring behaviour. Schmool \textit{et al.} \cite{8} have performed ferromagnetic resonance measurements on the same FePt/Fe interface system and have observed significant changes in the FMR spectra for RM and ES samples.

Exchange spring systems offer various potential technological applications, for example, in the form of permanent magnets \cite{3}, mag-MEMS \cite{6} and MRAM devices \cite{9}. 
In the current paper we consider the exchange spring system which corresponds to the FePt/Fe bilayer samples discussed above, in which the magnetocrystalline anisotropy of the FePt layer aligns its magnetisation in the direction perpendicular to the film plane. 

The paper is organised as follows: in the following section we discuss the basic theory and micromagnetic simulations used to study the system, in \ref{sec:Results} we present results and discussions of these simulations and in the final section we will give the conclusions of our study.

\section{Theoretical Model}\label{sec:model}

\subsection{Basic theoretical considerations}
The free energy of a magnetic system can be expressed in the most general form as the sum of various contributions: exchange energy, 
$E_{ex}$, demagnetising energy, $E_{dem}$, magnetocrystalline anisotropy energy, $E_{K}$ and the Zeeman energy, $E_{0}$. This can then be expressed as:
\begin{equation}
E_{total}=E_{ex}+E_{dem}+E_{0}+E_{K} \ \ .
\label{eq:totall}
\end{equation}

The equilibrium condition of the magnetic system can frequently then be evaluated by minimising this energy with respect to the orientation of the magnetisation for example. The dynamics of coherent magnetisation behaviour is typically described using the Landau - Lifshitz - Gilbert (LLG) equation and takes the form:

\begin{equation}
\frac{d\overrightarrow{M}}{dt}=-\gamma \,  \overrightarrow{M} \times  \overrightarrow{H}_{eff}+\frac{\alpha}{M_{s}} \left\{  \overrightarrow{M}\times \frac{\partial M}{\partial t}  \right\} \ \ ,
\label{eq:LLG}
\end{equation}
This equation describes the temporal evolution of the direction of the magnetisation vector due to the precessional motion induced by the magnetic torque associated with the effective applied field, $H_{eff}$ and its subsequent relaxation (viscous damping) which is controlled by the magnitude of the Gilbert damping parameter, $\alpha$. The effective magnetic field will have various contributions:
\begin{equation}
\overrightarrow{H}_{eff} =\overrightarrow{H}_{k}+\overrightarrow{H}_{0}+\overrightarrow{H}_{ex} - \overrightarrow{H}_{dem} \ \ ,
\end{equation}
and is related to the free energy of the system as described by the relation:
\begin{equation}
\vec{H}_{eff}=-\dfrac{\partial E_{total}}{\partial \vec{M}} \ \ .
\end{equation}
The boundary conditions of a magnetic film can have a significant influence on the dynamic response of the magnetisation to an excitation field or torque \cite{10}. This is because these boundary conditions are related to the surface and interface anisotropies and can effectively pin the magnetic spins at such magnetic discontinuities. For example, in a thin magnetic film with a magnetic field applied perpendicular to the plane standing spin waves can be excited, giving a resonance condition which is described by:

\begin{equation}
\omega_{n} =\gamma \left[ H_{0}- H_{k} -M_{0} +D\, k_{n}^2 \right]\ \ .
\label{eq:frequency}
\end{equation}
where $k_n$ is the spin wave wavevector of mode $n$. Restrictions imposed by the boundary conditions on the allowed values of the wavevectors can be expressed as \cite{11}:
\begin{equation}
\dfrac{k\, \left(p_1+p_2\right)}{p_1 \, p_2 -k^2}=\tan\left(kL\right) \ \ ,
\label{eq:valor_k}
\end{equation}
$p_1$ and $p_2$ are the pinning parameters which depend on the surface or interface energies and hence will be related to surface and interface anisotropies.

\subsection{Simulations}

In this work we concentrated on the study of the static and dynamic properties of a bilayer system where, for the most part, the interface has a rigid interface condition. This means that the interface spins at the Fe/FePt boundary are fixed in the perpendicular direction (which we take as the y-axis of the coordinate system). The dynamic response comes entirely from the Fe layer for our simulations (which we introduce in light of experimental results, which show that this is active in the frequency range studied \cite{8}). The simulations were performed by constructing a bilayer of FePt, with perpendicular anisotropy (in the L$1_{0}$ phase \cite{7, 12}), coupled to an Fe layer using the OOMMF (Object Oriented Micromagnetic Framework) \cite{13} software. The Fe film thickness was variable, in general, but for the dynamics we have maintained this constant, once we have predefined the thickness necessary to obtain an exchange spring systems, which corresponds to $25 \times 10^{-10}$ m. We note that the lateral dimensions were arbitrarily chosen and are much larger than the film thickness. The Fe layer is considered to have shape anisotropy and negligible magnetocrystalline anisotropy (as compared to the FePt layer). A Zeeman contribution will arise from the application of a magnetic field.

\begin{figure}[htpb]
\begin{center}
\includegraphics[width=9cm]{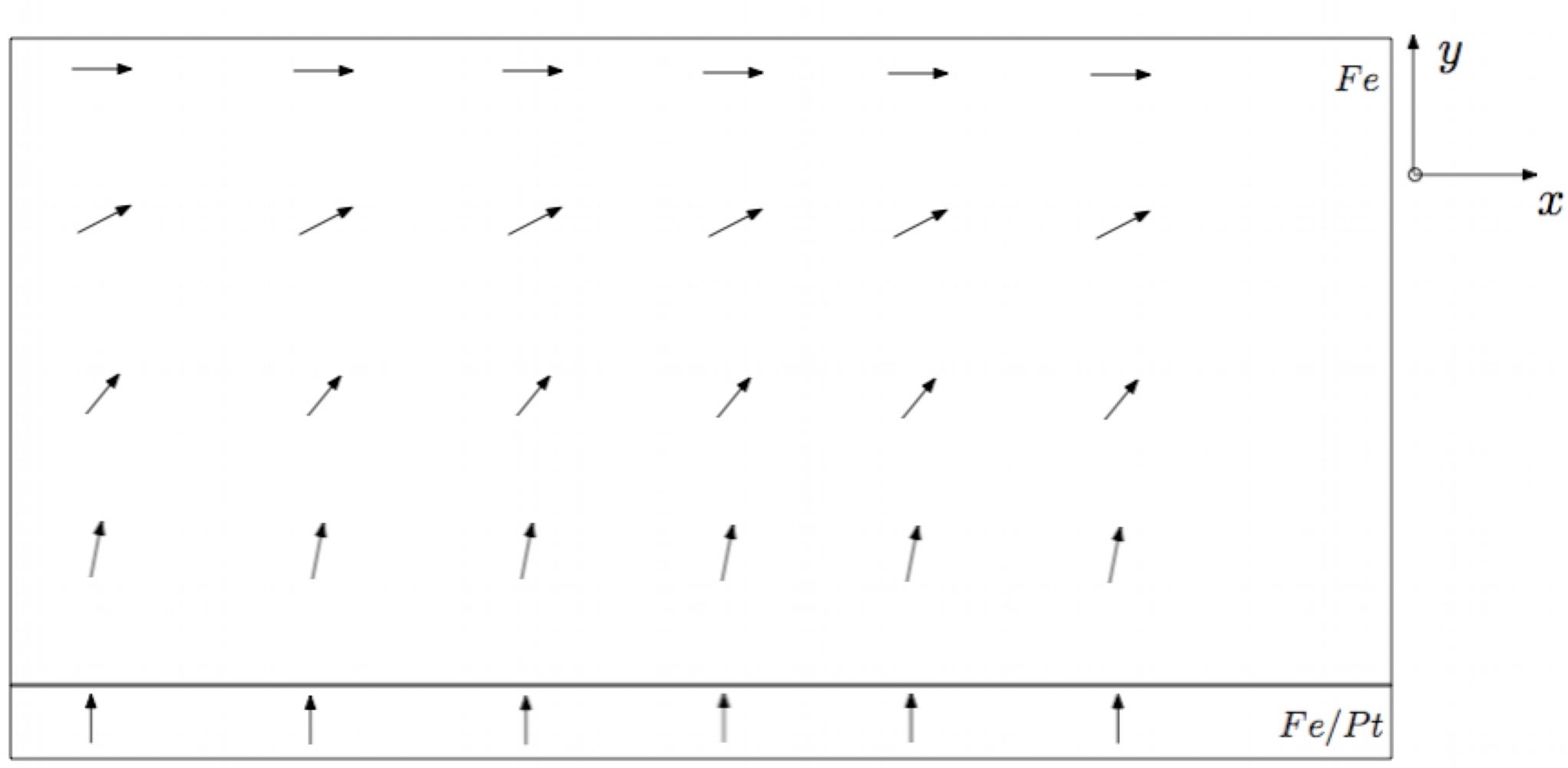}
\caption{Schematic representation of the position of the spins in the sample.}
\label{fig:spin_camada}
\end{center}
\end{figure}

The precision of the numerical results are governed by the micromagnetic cell parameters; $(10000, 1, 10000) \times 10^{-10}$ m. This essentially means that we are taking the cell to be one spin in the direction perpendicular to the film plane, where we measure the direction with respect to the xz plane orientation, as illustrated in fig \ref{fig:spin_camada}. The exchange constant between Fe/Fe and FePt/Fe spins is $1 \times 10^{-12}$ J/m, the uniaxial anisotropy in Fe/Pt layer is $4 \times 10^{6}$ J/m$^{3}$ (in y direction) \cite{d}, the damping coefficient is $0.01$ and the time step is $1 \times 10^{-14} s$. Finally the magnetisation saturation for Fe and Fe/Pt layers are $1700 \times 10^{3}$ and $690 \times 10^{3}$ A/m \cite{d}, respectively.

	In order to perform the dynamical calculations we need to obtain the equilibrium spin configuration, which is obtained by an energy minimisation routine without any static applied field. This is then taken as the initial condition for any subsequent dynamical calculations. The dynamics of interest in the Fe layer then arise due to the perturbation or excitation of the system upon the abrupt application of a magnetic field along some specific direction with respect to the film plane. The evolution of the magnetisation is then evaluated as a function of time, from which we perform a Fast Fourier Transform (FFT) to obtain the relevant frequency spectrum. We have used the z-component of the magnetisation for all calculations of the spin dynamics since it is transversal to the direction of the applied field, which is maintained in the x-y plane.

\section{Results and Discussion}\label{sec:Results}

\subsection{Static configuration}\label{Static}
\begin{figure}[ht]
\centering
\subfigure[Angle with respect to the $\widehat{y}$ direction as a function of the number of spins for a rigid interface.]{
\includegraphics[width=9cm]{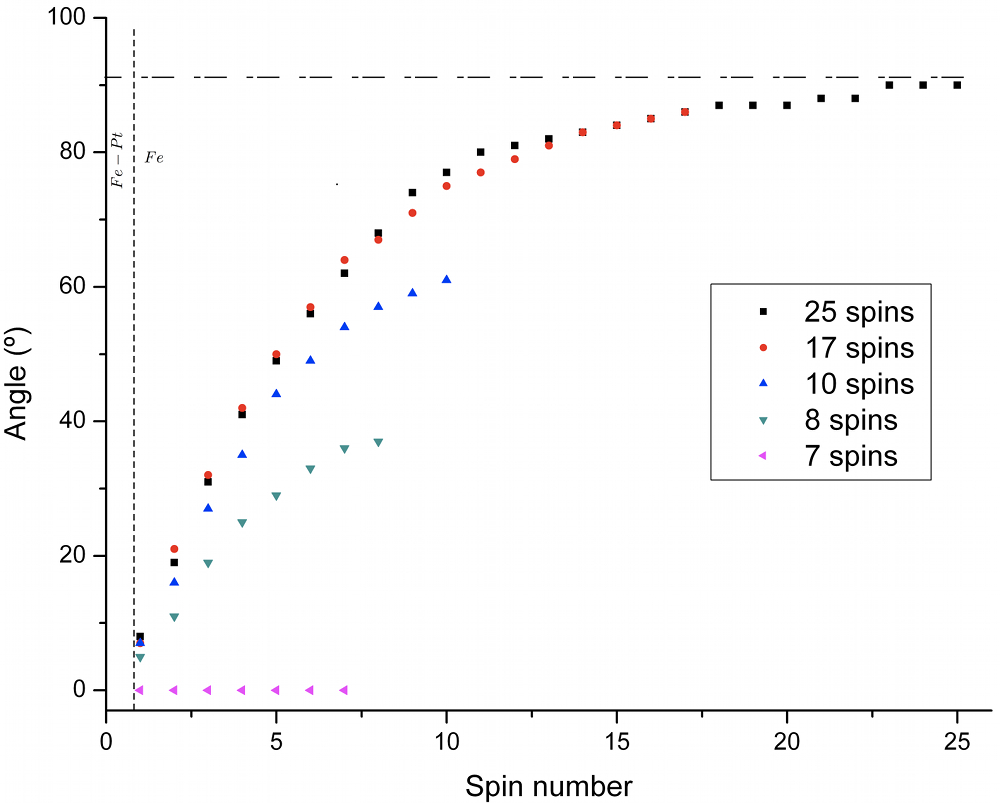}
\label{fig:static1}
}
\subfigure[ Angle with respect to the $\widehat{y}$ direction as a function of the number of spins for a 5 spins damped interface.]{
\includegraphics[width=9cm]{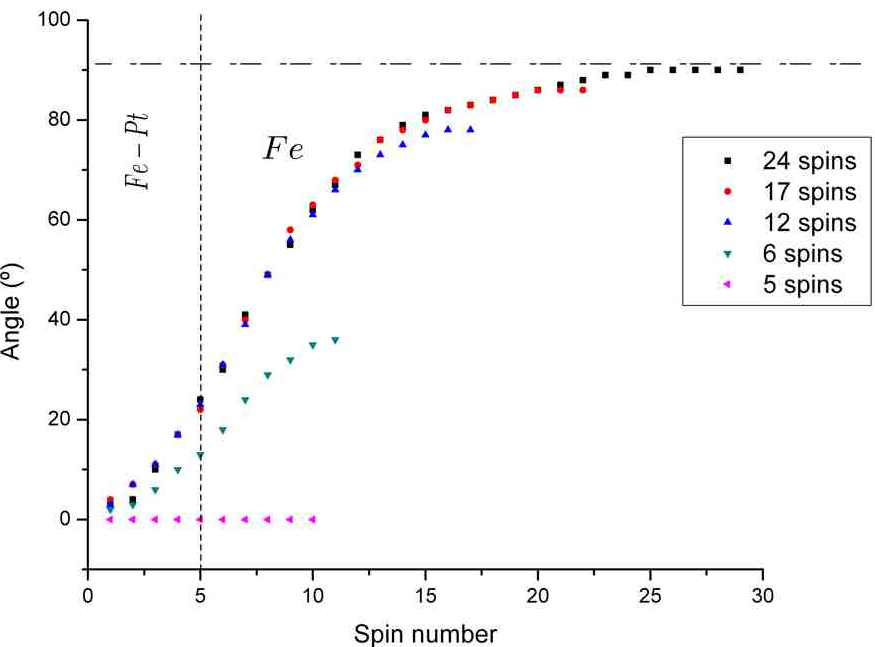}
\label{fig:static2}
}
\label{fig:estatic}
\caption[Dispersion relations]{\subref{fig:static1}, \subref{fig:static2} Show the spin angle as function of position in Fe layer. The first spin on each plot corresponds to the first unfixed spin in the chain.}
\end{figure}

The simulations were performed using the conjugate gradient method in the OOMMF package to evaluate the spin orientation as a function of its position in the Fe layer. We have varied the Fe layer thickness (number of spins) in order to find an optimum thickness for which to obtain an exchange spring (ES) system (i.e. a rotational variation across the film thickness). The results for the rigid interface (first Fe spin fixed in the perpendicular direction) are shown in figure \ref{fig:static1} . Here we note that for thicknesses corresponding to $7$ Fe spins, the system forms a rigid magnet (RM), whereby the Fe spins align along the FePt anisotropy (easy) axis as defined along the film normal. Adding one more spin we see that there is a significant relaxation to the spins in the Fe layer towards the in-plane direction. This relaxation further increases as further Fe spins are added. Almost full in-plane rotation (of the top Fe spin) occurs for around $25$ Fe spins. The transition from RM to ES is expected from energy considerations since at a certain thickness there are sufficient spins to allow the competition between the exchange coupling energy between the layers (causing an effective perpendicular anisotropy in the Fe layer) and the magnetostatic energy to be spread among the spins creating a domain wall. Full in-plane rotation can be seen as the construction of a $90^{\circ}$ domain wall (DW). Such a transition has been predicted, for example by Asti \textit{et al.} \cite{6} and observed experimentally by Casoli \textit{et al.} \cite{7}. In addition to considering the rigid interface conditions, we have performed the corresponding calculations for a relaxed interface. In this case we define the FePt layer to consist of $5$ spins. This means that the domain wall can now penetrate the hard magnetic layer, helping to further reduce the total energy of the system; the results are shown in figure \ref{fig:static2}. Once again a transition from RM to ES is observed, though the number of Fe spins is reduced compared to the rigid interface condition. This can be expected since the extension of the DW into the FePt layer helps shift some of the energy from the Fe layer to that of the FePt. The static configuration is a sensitive function of the number of spins in both the Fe and FePt layers. We have performed extensive studies of such conditions and will be reported elsewhere \cite{14}. 
For our purposes, we have chosen to study the Fe layer consisting of $25$ spins using rigid interface conditions. This offers a system in which an almost complete $90^{\circ}$ DW wall exists. We now proceed to study the magnetisation dynamics in this system.

\subsection{Magnetisation dynamics}

\begin{figure}[ht]
\centering
\subfigure[ FFT for magnetic field of $1$ T and $20^{\circ}$.]{
\includegraphics[width=8cm]{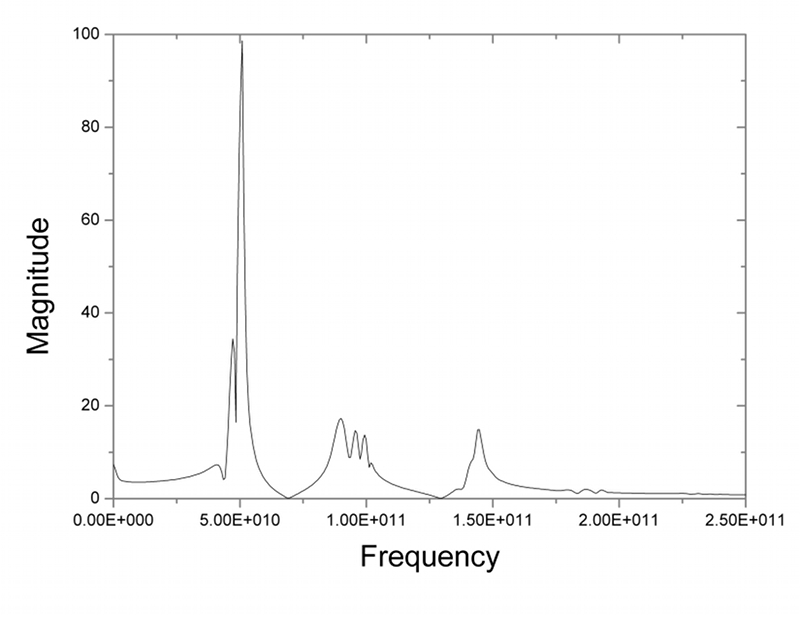}
\label{fig:positivo}
}
\subfigure[ FFT for magnetic field of $1$ T and $-20^{\circ}$.]{
\includegraphics[width=8cm]{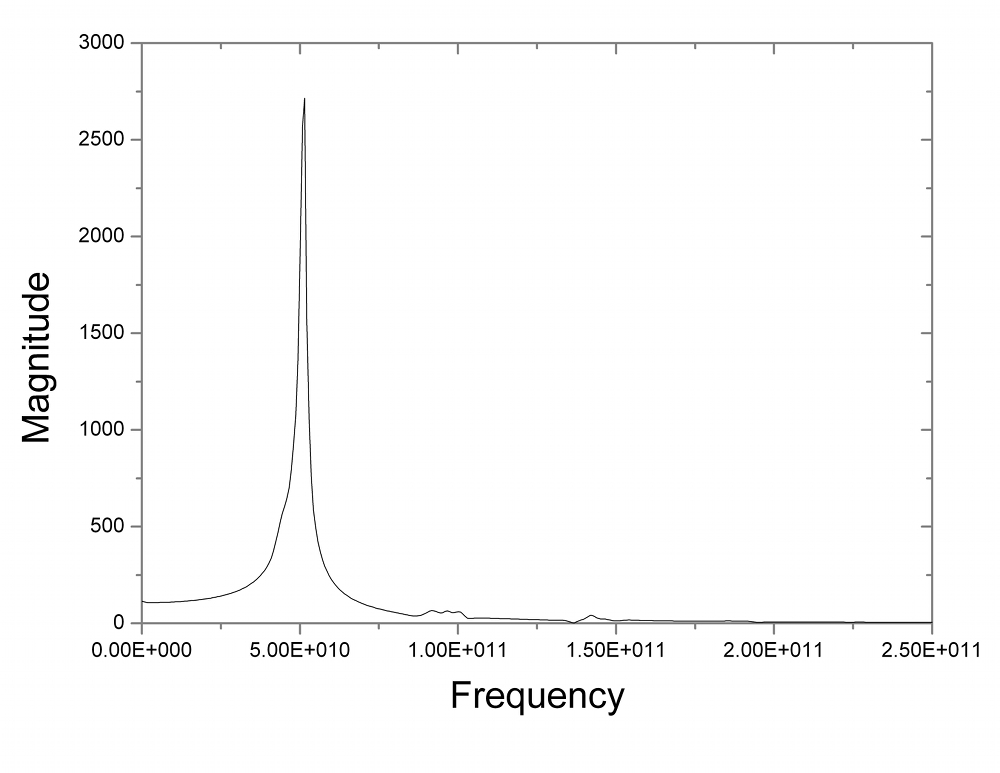}
\label{fig:negativo}
}
\label{fig:angle1}
\caption[FFT spectrums]{\subref{fig:positivo}, \subref{fig:negativo} show the FFT spectrum of the Fe layer.}
\end{figure}

To study the magnetisation response to an abruptly applied magnetic field along some specific direction, we use the LLG, which is solved using the Runge-Kutta method, also in the OOMMF software. To understand the importance of material parameters and response of the system, we have made a series of simulations with applied magnetic fields of different strengths applied in various directions. The perpendicular anisotropy of the FePt layer provides the system with a unidirectional character which will be transmitted to the Fe layer via the exchange coupling between the layers. Such properties will be expected to influence the dynamic properties of the Fe layer. This can be illustrated, for example, in the spectra obtained for the $+20$ and $-20^{\circ}$ directions, which are shown in figure \ref{fig:positivo} and \ref{fig:negativo}. 

\begin{figure}[ht]
\centering
\subfigure[ Plot of the resonance frequency as a function of the applied magnetic field for positive angles (measured in the first quadrant).]{
\includegraphics[width=9cm]{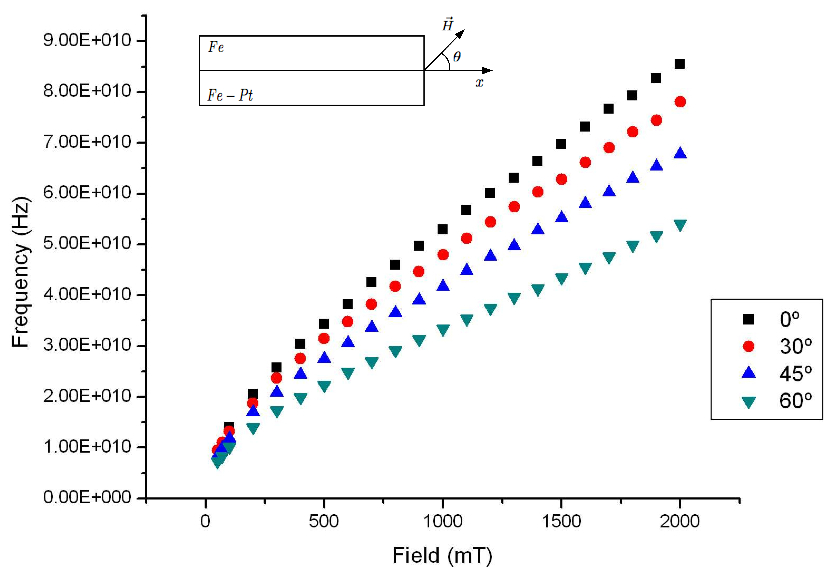}
\label{fig:dispersaopositiva}
}
\subfigure[ Plot of the resonance frequency as a function of the applied magnetic field for negative angles (measured in the fourth quadrant).]{
\includegraphics[width=9cm]{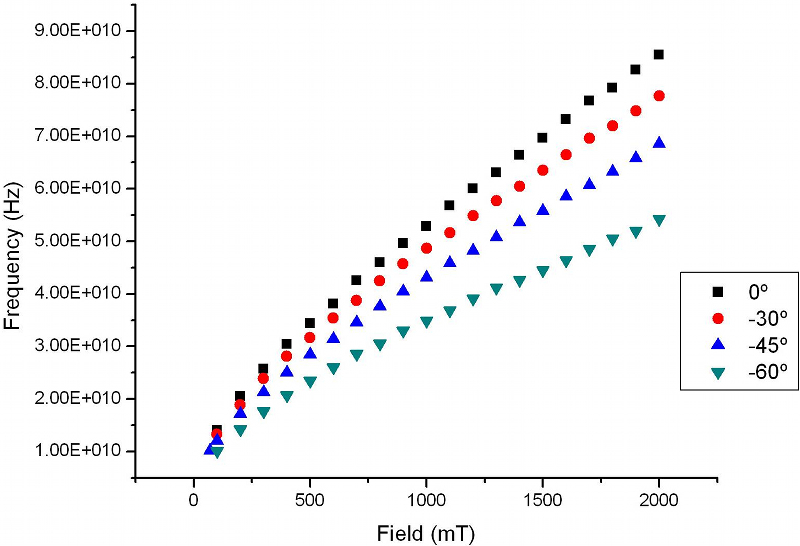}
\label{fig:dispersaonegativa}
}
\label{fig:dispersao9}
\caption[Dispersion relations]{\subref{fig:dispersaopositiva}, \subref{fig:dispersaonegativa} depict the spectrum of frequencies as a function of the applied magnetic field for several angles.}
\end{figure}

We note that the principle position (frequency) of the peaks in the spectra are the same; however, the relative intensities are very different - the principal peak in figure 3(b) shows a higher intensity than in figure 3(a). Therefore this camouflages the importance of the other relevant peaks (frequency modes). Some of the principle characteristics of the dynamical properties can be elucidated from the study of the frequency  field behaviour, which we have done as a function of the direction of the applied magnetic field. In \ref{fig:dispersaopositiva} and \ref{fig:dispersaonegativa} we show the dispersion relations of the dominant resonance frequency for various angles of the applied field. (We note that the $0^{\circ}$ direction corresponds to the field applied in the plane of the film.) While the positive and negative directions appear to give very similar results, there are some important differences which are related to the unidirectional character of the system. The variation in each case appears as for a system with almost vanishing magnetic anisotropy, which is clearly not the case.

\begin{figure}[ht]
\centering
\subfigure[ Resonance frequency as a function of the angle for fixed applied magnetic field.]{
\includegraphics[width=7cm]{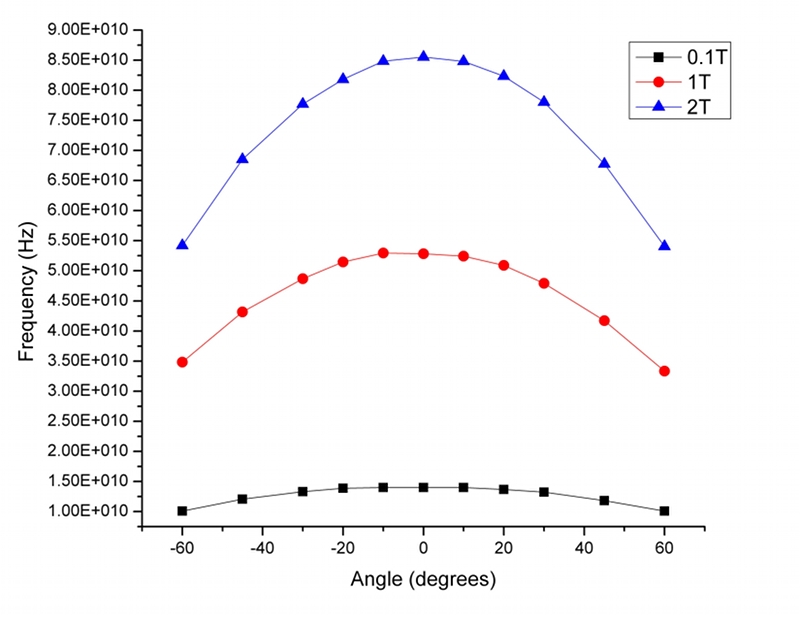}
\label{fig:angle}
}
\subfigure[ Resonance frequency per applied magnetic field as a function of the angle.]{
\includegraphics[width=9cm]{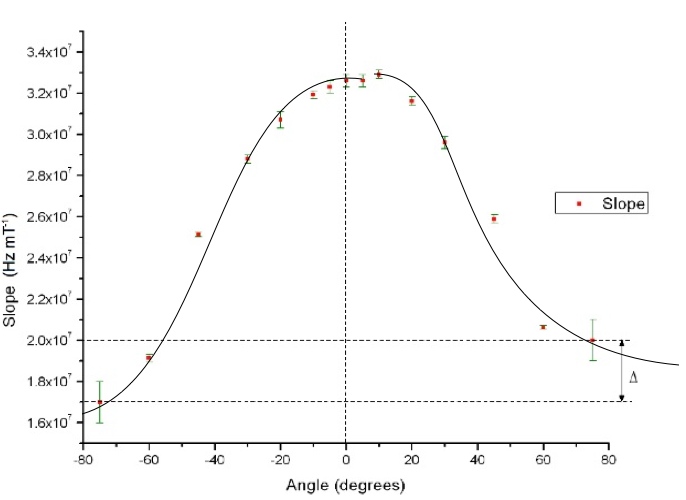}
\label{fig:slope}
}
\caption[Stuff]{\subref{fig:angle}, \subref{fig:slope} show the symmetry properties of the system.}
\label{fig:aiu}
\end{figure}

In a system with a well defined anisotropy axis, the frequency  field behaviour has a characteristic shape and depends on the type of magnetic anisotropy; uniaxial, cubic, etc. For example, in the case of uniaxial anisotropy, with the field applied along the hard axis, the dispersion relation shows a marked cusp with a minimum corresponding to the anisotropy field, $H_{K}$, of the system. While with the field applied along the easy axis the intercept with the frequency axis occurs to a resonance condition in which a field corresponding to the anisotropy field is applied; i.e. $\omega = \gamma H_{K}$, see for example Vonsovskii \cite{15}. In the present case we see no well defined axis of anisotropy. This is because in our case this does not exist, since the directions of the spins in the Fe layer vary as a function of position across the thickness of the film. This will mean that we cannot, strictly speaking, treat the system in the usual manner, despite the usual asymptotic behaviour still being apparent \cite{16}. In figure \ref{fig:angle} we illustrate the resonance frequency as a function of angle (of the applied magnetic field) for $0.1$, $1$ and $2$ T. 
This shows a reasonably symmetric variation about the x-axis. However, a closer inspection reveals that for negative angles the frequencies are higher. 

A more sensitive display of this asymmetry is shown in the variation of the asymptotic slope of the frequency  field data as a function of angle, see figure \ref{fig:slope}. The disparity between positive and negative directions becomes more marked for larger angles. In fact it would appear that there is a discontinuity in the variation in the region of $5^{\circ}$. This asymmetry and discontinuity, as well as the asymmetry in the angular dependence of the frequency are related to the unidirectional behaviour transmitted to the Fe layer from the FePt layer via the exchange interaction. Added to this the rigid boundary conditions will be expected to accentuate this behaviour.

\begin{figure}[htpb]
\begin{center}
\includegraphics[width=8cm]{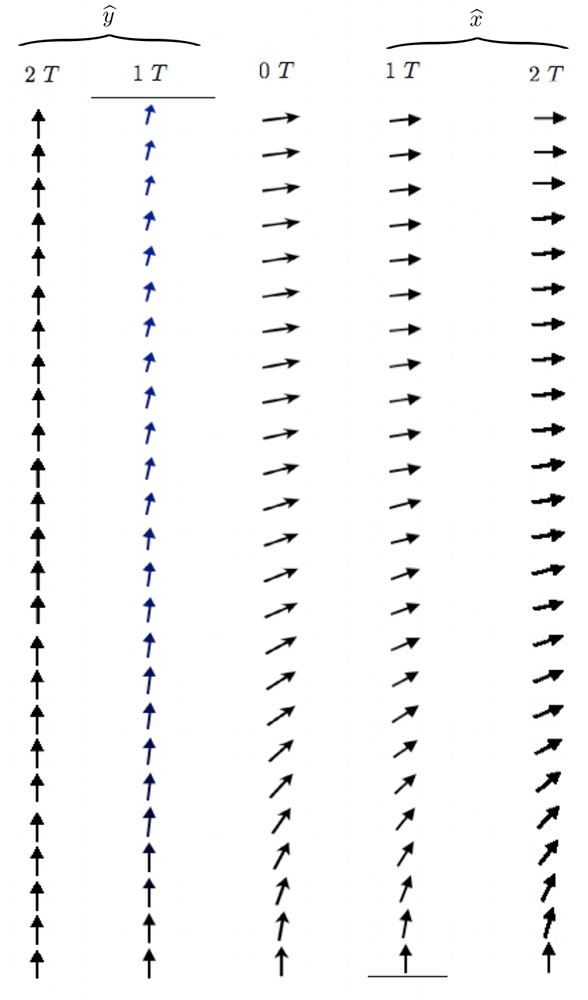}
\caption{Initial spins configuration ($0$ T); final spin configuration in $\widehat{y}$ ($90^{\circ}$) and $\widehat{x}$ ($0^{\circ}$) direction. The first spin, in the botton of each chain, is the fixed spin from the FePt layer. }
\label{fig:configuration}
\end{center}
\end{figure}

To gain a fuller understanding of the dynamical response of the system to the abruptly applied fields in the simulation, it is instructive to consider the initial and final configurations; i.e. that of the initial state without an applied field (which will be the same for all simulations) and the final equilibrium state to which the system will relax. In figure \ref{fig:configuration} we show the spin configurations for specific cases: the $0$ T case corresponds to the initial spin configuration used for each simulation. 
We also show the final state configurations for $1$ and $2$ T applied along the film plane ($x$-axis) and along the normal direction to the film plane (y-axis). In all cases the lower (Fe) spin is always in the perpendicular direction and arises due to the fixed (or rigid) boundary condition that we have imposed. In the case of a $2$ T field applied along the film normal, all spins align along the field giving a saturated state. It will be noted that $1$ T is not sufficient to push the DW out of the film and a small rotation is still evident. Applying the field in the plane of the film effectively squeezes the $90^{\circ}$ DW, reducing its thickness. 

The transition from initial to final state will define the dynamics of the magnetisation process via the precessional motion of each spin. There will be two main factors which will govern this motion; the effective field; $H_{eff}$, for each spin, which will depend on its position within the layer, and the overall angle difference between the initial and final states. The fact that the effective field varies as a function of position can be seen by the equilibrium spin distribution across the Fe layer. The difference between initial and final state will govern in large part the amplitude of the motion of each spin in its movement; 
\begin{figure}[htpb]
\centering
\subfigure[ Variation of the angle between initial and final configurations as a function of the position of the spin in the sample in the $\widehat{x}$ direction for $1$ T and $2$ T.]{
\includegraphics[width=8.5cm]{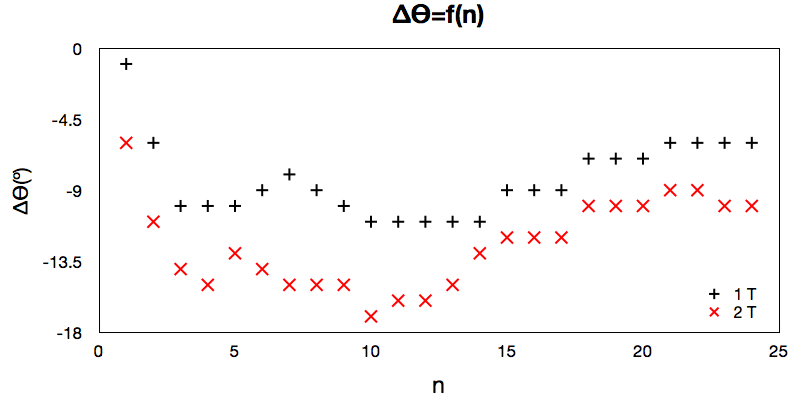}
\label{fig:delta_x}
}
\subfigure[ Variation of the angle between initial and final configurations as a function of the position of the spin in the sample in the $\widehat{y}$ direction for $1$ T and $2$ T.]{
\includegraphics[width=8.5cm]{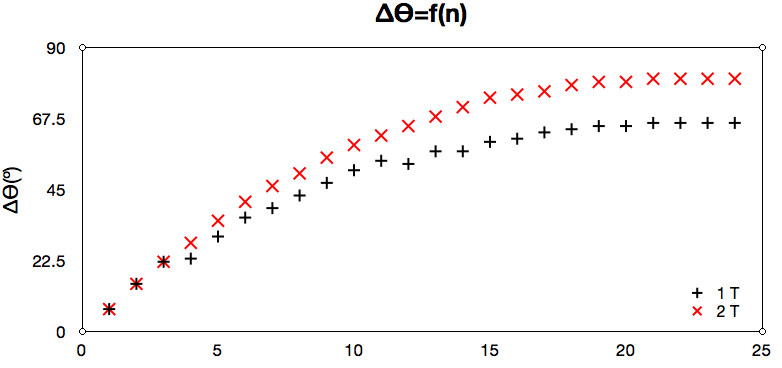}
\label{fig:delta_y}
}

\caption[Stuff23]{\subref{fig:delta_x}, \subref{fig:delta_y} show the relative deviations from the initial angle. The fluctuations observed in figure 7(a) are due to the error in determining the angle. Only the line of tendency should be considered and not the particular value of each point in the curve.}
\label{fig:airf}
\end{figure} 
figure \ref{fig:airf} shows these variations for the parallel and perpendicular fields illustrated in figure \ref{fig:configuration}. 
It will be noted that for in-plane fields the variation is negative, while for out of plane fields this will be positive (due to the convention we have chosen). This will therefore determine the character of each spectrum and can be related to the resonance frequency and the relative intensities of the various peaks in the spectra. In addition to the amplitude or precessional angle of motion, the intensity of a particular peak will also depend on the number of spins which are involved in that motion. To illustrate this interpretation we can consider some specific cases. Let us first consider the field applied in the plane of the sample (along the x-axis). The initial and final spin configurations are illustrated in figure \ref{fig:configuration}. The first thing to note is that there is only a small change in the configuration itself, with the largest changes occurring at around the $10$th spin from the interface between Fe and FePt. For Fe layers of different thicknesses this will be different. We also note that the differences should increase with the strength of the applied field; see figure \ref{fig:delta_x}. 
When the magnetic field is applied along the film normal we notice that the character of the spectrum becomes much more regular; the spectrum for a $2$ T field applied in the perpendicular direction is shown in figure \ref{fig:spec}. 
\begin{figure}[htpb]
\begin{center}
\includegraphics[width=8cm]{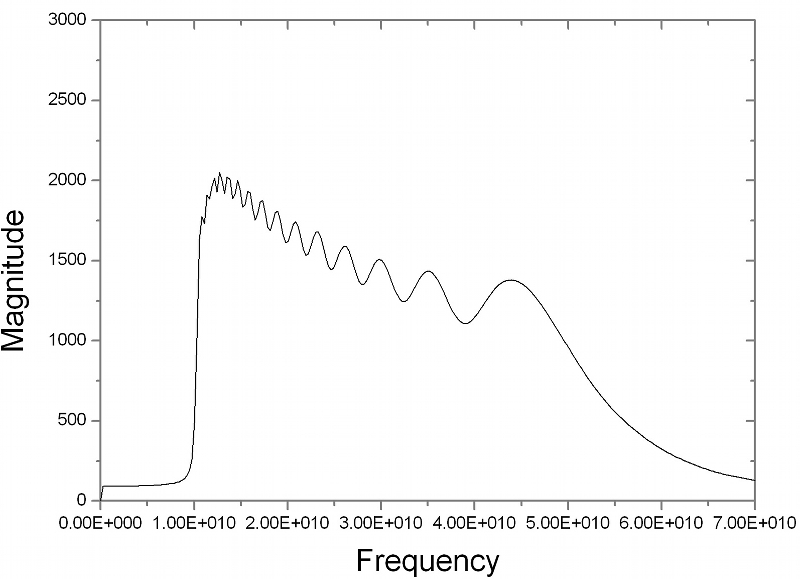}
\caption{Spectrum of frequencies obtained by FFT for an applied magnetic field of $2$ T and $90¼$ (with respect to the $\widehat{y}$ direction).}
\label{fig:spec}
\end{center}
\end{figure}
Similar spectra are obtained for other field values. In this case the differences from initial to final state are much greater, and increase as we move from the interface to the outer surface of the Fe layer, as shown in figure \ref{fig:delta_y}. In traditional magnetodynamic measurements performed by ferromagnetic resonance (FMR), for example, multi-peaked spectra are usually associated with the existence of standing spin wave resonance modes which arise from the magnetic confinement in the perpendicular direction and the specific boundary conditions which apply. Such considerations lead to a resonance equation for the resonance frequency in function the applied magnetic field and the spin wave wave-vector of the form of the Kittel equation \cite{17} - see equation \ref{eq:frequency}.

\begin{figure}[htpb]
\begin{center}
\includegraphics[width=9cm]{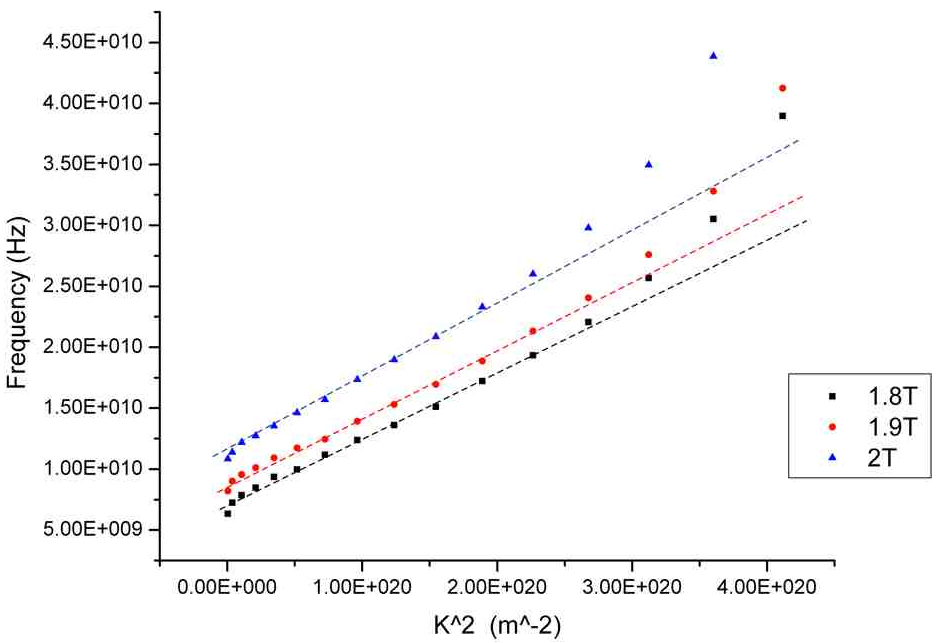}
\caption{Frequency as a function of the squared wavenumber for different magnetic fields using the Kittel theory for spin waves.}
\label{fig:spin}
\end{center}
\end{figure}

If we consider that the spins at the lower interface are fixed in the perpendicular direction, while those on the outer surface are free to move (as in the bulk), we can easily arrive at the allowed values for the spin wave modes as given by:

\begin{equation}
k=\dfrac{\left(2n-1\right)}{2L} \, \pi \ \ \ \ \ , \ \  n= 1,2,3... \ ,
\label{eq:def_k}
\end{equation}
where $n$ denotes the mode number and $L$ the thickness of the Fe layer. In this case both odd and even modes should be allowed since the boundary conditions are non-symmetric. A plot of frequency against the square of the wave-vector should yield a linear variation - according to equation \ref{eq:frequency}. Using our data we have made such plots for various applied fields along the perpendicular direction; these are shown in figure \ref{fig:spin}. We note that while much of the data fit on linear portions, the extremal values are clearly not in agreement. Therefore we believe that the various peaks in the spectra do not arise from the excitation of standing spin wave modes. Rather, we think that the existence of the multi-peaked spectra is due to the different local conditions (local effective fields) of each spin, there will be a distribution of resonance conditions which are sufficiently spread out as to be distinguishable resonances. These local effective fields will reflect the differing conditions and the local environment, where a local energy minimum is encountered, giving rise to the direction of the spin at that position. The intensity distribution also reflects the initial to final state directions, which in much of the layer are quite large (see figures \ref{fig:configuration} and \ref{fig:aiu}). This will mean that the initial precessional angles will be large and hence give increased intensities, which are proportional to the transversal component of the magnetisation.

\section{Conclusions}\label{sec:summary}
We have made a detailed study of the model system comprised of a hard magnetic layer (FePt) with perpendicular anisotropy exchange coupled to a soft magnetic layer (Fe). The first step to the study of the magnetisation dynamics in this system concerns the determination of the equilibrium condition. This is achieved by minimising the free energy density of the system and obtaining the orientation of the individual local spin orientations. We have performed such calculations for the system with varying Fe thicknesses. Since we are interested in studying the exchange spring system, we use these calculations to choose the Fe thickness we require; i.e. where there is significant rotation of the magnetic moment through the film. We see the transition from the RM to ES occurs for a system with a thickness corresponding to $6$ and $7$ atomic spins for rigid and free boundary conditions, respectively. This corresponds to a thickness of around $0.86$ and $1.0$ nm and compares reasonably well with the calculations of Asti \textit{et al.} \cite{6}, who obtain a value of between $0.6$ and $0.8$ nm, depending on the thickness of the FePt layer. 

The dynamics of the system - in response to an abruptly applied magnetic field - were studied as a function of the strength and direction of the applied field. This allowed us to obtain the frequency spectrum and therefore construct the dispersion relation. We have considered only the dynamics of the Fe layer, since this should be separate from the FePt dynamic response due to its elevated anisotropy, and is supported by FMR measurements which only detect Fe resonance features at low frequencies \cite{8}. We note that the lack of a well defined anisotropy axis in the Fe layer means that the angular variation of the frequency  field characteristic differ from that of a normal film, and always pass through the origin. The high field asymptotic gradient varies with direction, whose angular dependence is related to the exchange coupling with the hard layer. A small asymmetry is noted around the x-axis which is due to the unidirectional character of this exchange coupling. For magnetic field applied in the perpendicular direction we obtain a rich frequency spectrum which can be understood in terms of the variation of the local effective magnetic field across the Fe layer.

As an aside, it is worth noting that the OOMMF software provides sufficient resolution to allow us to treat the system as atomic layers of spins in the direction perpendicular to the film plane. This is supported, for example, by the fact that we were able to reproduce the RM-ES transition, and is supported by experimental measurements \cite{7} and other calculations \cite{14}.

\vspace*{0.5cm}
{\bf Acknowledgements}. We thank OOMMF mailing list for insightful discussions.

\appendix

\end{document}